\documentclass[%
 reprint,showpacs,
 amsmath,amssymb,
 aps,
]{revtex4-1}

\usepackage{graphicx}
\usepackage{dcolumn}
\usepackage{bm}
\usepackage{amssymb,amsfonts,amsmath,natbib}
\usepackage{hyperref}

\newcommand{\etal}{{\it et al.}}

\newcommand{\rmi}{\mathrm{i}}

\newcommand{\Fra}[2]{\displaystyle \frac{ #1}{ #2}}

\newcommand{\dg}{^\circ}
\newcommand{\de}{\,\mathrm{d}}

\newcommand{\Dpar}[2]{\displaystyle \Fra{\partial #1}{\partial #2}}

\newcommand{\Lapl}[1]{\displaystyle \boldsymbol{\nabla}^2 #1}

\newcommand{\Grad}[1]{\displaystyle \boldsymbol{\nabla} #1}

\begin{document}
\title{Helicity conservation under quantum reconnection of vortex rings}
\author{Simone Zuccher}%
\email{simone.zuccher@univr.it}%
\affiliation{Department of Computer Science, U. Verona,
Ca' Vignal 2, Strada Le Grazie 15, 37134 Verona, Italy}%
\author{Renzo L. Ricca}%
\email{renzo.ricca@unimib.it}
\affiliation{Department of Mathematics \& Applications,  
U. Milano-Bicocca, Via Cozzi 55, 20125 Milano, Italy}%
\date{\today}
\begin{abstract}
Here we show that under quantum reconnection, simulated by using the three-dimensional 
Gross-Pitaevskii equation, self-helicity of a system of two interacting vortex rings remains conserved. By resolving the fine structure of the vortex cores, we demonstrate that total length of the vortex system reaches a maximum at the reconnection time, while both writhe helicity and twist helicity remain separately unchanged throughout the process. Self-helicity is computed by two independent methods, and topological information is based on the extraction and analysis of geometric quantities such as writhe, total torsion and intrinsic twist of the reconnecting vortex rings.
\end{abstract}

\maketitle

\section{\label{sec:intro}Introduction}
\textit{Background.}---Reconnection of coherent structures play a 
fundamental r\^ ole in many areas of science.
Examples include vortices in classical fluid flows~\cite{KT1994,HD2011}, 
quantum vortex filaments in superfluid Helium~\cite{PFL2010,ZCBB12}, 
magnetic flux tubes in plasma physics~\cite{LF96,PF2000}, phase transitions in mesoscopic 
physics~\cite{L99}, macromolecules in DNA biology~\cite{VCP98}.
Here we focus on a single reconnection event, that characterizes  superfluid quantum 
turbulence~\cite{VIN2008,BSS2014}, by analyzing dynamical,  
geometric and topological properties that are relevant also 
in classical viscous fluids~\cite{HD2011}, where similar features 
such as time asymmetry~\cite{ZCBB12}, helicity transfers, 
randomization of the velocity field and energy cascades~\cite{K11} 
are important.

In recent months a number of remarkable results based on experimental 
observations~\cite{SKP14}, mathematical analysis~\cite{LRS15} and theoretical and numerical 
work~\cite{KM14} have provided contradictory 
information as regards helicity transfer through reconnection. 
On one hand laboratory experiments on the production and evolution of vortex knots in water show~\cite{SKP14} 
that the centerline helicity of a vortex filament  
remains essentially conserved throughout the spontaneous  reconnection 
of the interacting vortices. This result is mirrored by the mathematical analysis of conservation 
of writhe and total torsion (for definitions, see~Sec.\ref{sec:topquant} here below) 
under the assumption of anti-parallel reconnection of the interacting strands~\cite{LRS15}. 
On the other hand recent
numerical results~\cite{KM14}, based on a linearized model of interacting 
Burgers-type vortices brought together by an ambient irrotational strain field, show that 
the initial helicity associated with the skewed geometry is eliminated during the process. 
This apparent contradiction motivates further the present study. 

In this paper we carry out a simulation of 
the interaction and reconnection of a single pair of identical quantum vortex rings. The 
evolution is governed by the three-dimensional Gross-Pitaevskii equation (GPE), with the aim to 
reproduce and analyze in the GPE context the fine details of the prototype
reconnection event as studied in~\cite{KM14}. By resolving the 
fine structure of the vortex cores, we monitor all the relevant dynamical, geometric and 
topological features of the reconnection process. Consistently with current simulations (see, 
for example~\cite{K11}), the peak in the normalized total length of the vortex system, given 
by an initial stretching process followed by its marked decay, 
is taken as signature of the reconnection event, providing a precise benchmark for the 
diagnostics of the mathematical and physical properties associated with the reconnection event.

\textit{Governing equations.}---Direct numerical simulation of the reconnecting quantum vortex 
rings is done by using the 3D Gross-Pitaevskii equation~(GPE)
\begin{equation}\label{eq:GPE}
  \frac{\partial \psi}{\partial t} = 
  \Fra \rmi 2 \Lapl \psi + \Fra \rmi 2(1 - |\psi|^2)\psi \ ,
\end{equation}
with background density $\rho_b=1$. Through the Madelung transformation 
$\psi = \sqrt\rho\, \mathrm{exp}(\rmi \theta)$, eq. \eqref{eq:GPE} admits decomposition into 
two equations that can be interpreted in classical fluid dynamical terms, i.e. the continuity 
equation and the momentum equation, given by 
\begin{eqnarray}
\label{eq:GPEinNScont}
  &&\Dpar{\rho}{t}+\Dpar{(\rho u_j)}{x_j}=0\ ,\\
\label{eq:GPEinNSmom}
  &&\rho \left(\Dpar{u_i}{t}+u_j\Dpar{u_i}{u_j} \right) =
  -\Dpar{p}{x_i} + \Dpar{\tau_{ij}}{x_j}\ ,
\end{eqnarray}
where $\rho=|\psi|^2$ denotes fluid density, ${\boldsymbol u}=\Grad{\theta}$ velocity, 
$p=\frac{\rho^2}{4}$ pressure, and
$\tau_{ij}=\frac 1 4 \rho\frac{\partial^2\ln\rho}{\partial x_i \partial x_j}$ the so-called 
quantum stress ($i,j=1,2,3$). 
Defects in the wave function $\psi$ represent infinitesimally thin vortices of constant 
circulation $\Gamma=\oint{\boldsymbol u}\cdot\!\de{\boldsymbol s}=2\pi$ of  healing length $\xi=1$. 
It is well known that GPE conserves mass, given by  
$M=\int|\psi|^2\de{\boldsymbol x}^3$, and the hamiltonian $E=K+I$, where
\begin{equation}\label{eq:ham}
  K=\frac12\int\Grad\psi\cdot\Grad\psi^*\de{\boldsymbol x}^3\ ,\  
  I=\frac14\int(1-|\psi|^2)^2\de{\boldsymbol x}^3\ ,
\end{equation}
denote respectively the kinetic ($K$) and interaction ($I$) energy of the system ($\psi^*$ 
being the complex conjugate of 
$\psi$). The term $\tau_{ij}$, negligible compared to the pressure 
term at length scales much larger than the healing length $\xi=1$, 
is expected to be key to vortex reconnection~\cite{ZCBB12}, and at scales larger than the vortex core, GPE 
in the form of eqs. \eqref{eq:GPEinNScont} and \eqref{eq:GPEinNSmom} reduces to the classical 
compressible Euler equations. 

\begin{figure}
\centering
\includegraphics[scale=0.18]{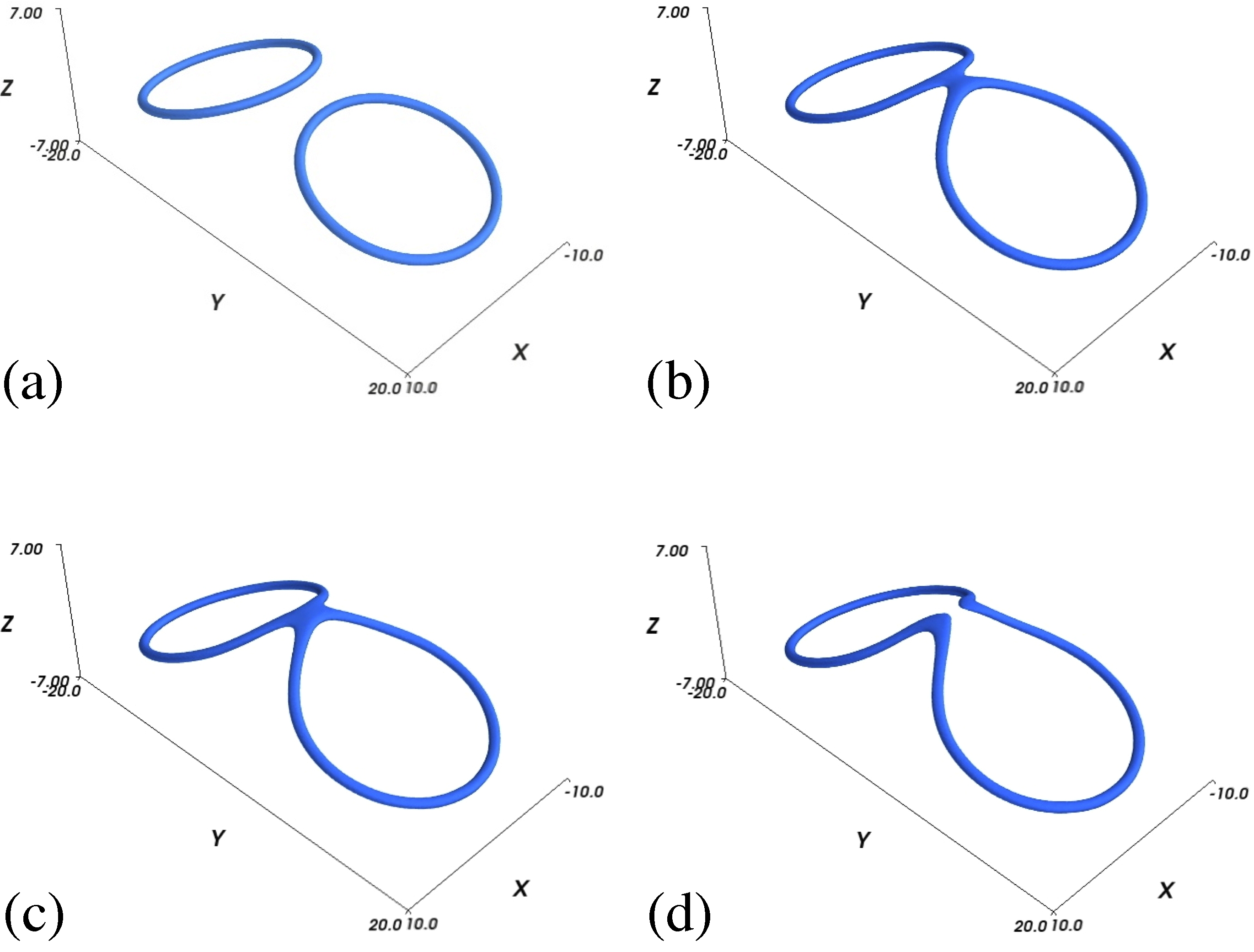}
\caption{\label{FIGURE1}
Time-evolution of interaction and reconnection of two quantum vortex rings;
isosurfaces of $\rho=0.06$. (a) $t=0$, (b) $t=10$,
(c) reconnection time $t=t^*= 11.81$, $t=14$.}
\end{figure}

\textit{Helicity and self-linking number.}---A fundamental quantity of 
topological fluid mechanics is kinetic helicity, defined by~\cite{S1992}  
\begin{equation}\label{eq:Hdef}
  H=\int{\boldsymbol u}\cdot{\boldsymbol\omega}\,\de{\boldsymbol x}^3\ ,
\end{equation}
where ${\boldsymbol\omega}=\nabla\times{\boldsymbol u}$ is vorticity and 
the integral is extended over the vorticity volume. $H$ is known to be an invariant of ideal fluid 
flows and in ideal conditions it admits a topological interpretation in terms of linking 
number~\cite{MOF69}. For a pair of linked vortex rings $V_1$ and $V_2$, centred respectively on 
curves $C_1$ and $C_2$ and of equal circulation $\Gamma$, eq. \eqref{eq:Hdef} can be written 
as~\cite{RM92,MR92}
\begin{equation}\label{eq:Hlinking}
  H(V_1,V_2)=\Gamma^2[SL(V_1)+SL(V_2)+2Lk(C_1,C_2)]\ ,
\end{equation}
where $H(V_1,V_2)$ is the total helicity of the system, $SL(V_i)$ is the 
(C\u alug\u areanu-White) self-linking number of $V_k$ $(k=1,2)$ and 
$Lk(C_1,C_2)$ is the (Gauss) linking number of $C_1$ and $C_2$. Note that 
if the pair of rings are unlinked (as in our case, cf. Figure~\ref{FIGURE1}a), 
then $Lk(C_1,C_2)=0$ and \eqref{eq:Hlinking} can be further 
simplified to 
\begin{equation}\label{eq:Hlinking1}
  H(V_1,V_2)=\Gamma^2[SL(V_1)+SL(V_2)]\ .
\end{equation}
In general the self-linking number $SL$, can be decomposed 
into global geometric quantities, and one can show~\cite{Calu61,White69} that 
$SL(V_k)=Wr(C_k)+T(C_k)+N(R_k)$,
where writhing number $Wr(C_k)$, total torsion $T(C_k)$ and intrinsic twist $N(R_k)$ 
are quantities that depend solely on the shape of the vortex centerline $C_k$ and 
ribbon $R_k$ (for definitions see~\cite{MR92, LRS15} and Sec.~\ref{sec:topquant} here below).

\section{\label{sec:numerics}Numerics and initial conditions}
The numerical code used for the simulation is described in~\cite{ZCBB12}.
It is based on a second-order Strang splitting method in time, and Fourier 
decomposition in space. Hence, boundary conditions must be periodic; for 
non-periodic directions the computational domain is doubled and ``mirror'' vortex 
rings are introduced in the doubled domain, as was done in~\cite{KL1993}.
The method conserves mass exactly.

\textit{Initial conditions.}---A pair of vortex rings is set at the center 
of the numerical box. While this particular setting provides a good comparative 
test for the physics of vortex reconnection~\cite{WTZ2014}, it helps to avoid
difficulties associated with the numerical implementation of boundary conditions 
and the topological complexity implied by periodic conditions, while offering a 
realistic match to simulate the event studied in~\cite{KM14}.

At time $t=0$ the two rings are centered in $(0;\pm 10;0)$, have radius 
$R_0=8$ and are mutually inclined by an angle $\alpha=\pm \pi/10$ 
with respect to the horizontal plane (see Figure~\ref{FIGURE1}a).
The computational domain is $[-20;20] \times [-30;30] \times [-20;20]$.
In order to have fine spatial and temporal resolution of the vortex core and of 
the reconnection process, we have used $\Delta x=\Delta y=\Delta z= \xi/6$ 
(i.e. the number of points is $240 \times  360 \times 240$) and $\Delta t=1/80=0.0125$.

At each point $Q$ on the vortex ring we place a Frenet triad 
$\{\mathbf{\hat t},\mathbf{\hat n},\mathbf{\hat b}\}$ given by the local unit tangent, normal 
and binormal to the vortex centerline (no inflexion points emerge during the simulation). For 
each grid point $P$ in the numerical domain we seek the 
nearest point $Q$ on the vortex line so that $\overrightarrow{QP}$ identifies 
the distance of $P$ from the vortex. Thus, $\overrightarrow{QP}$ is locally orthogonal to the 
vortex, in the plane defined by $\mathbf{\hat n}$ and 
$\mathbf{\hat b}$ at $Q$. In this plane $P$ has polar coordinates $(r,\theta)$ centred on $Q$, 
where $r=\overline{QP}$ and $\theta$ is the angle between 
$\overrightarrow{QP}$ and $\mathbf{\hat n}$.

Each vortex contributes to the initial condition with a density
distribution ${\rho_0}_k$ given by the Pad\'e approximation~\cite{BER2004}
${\rho_0}_k = \left(\frac{11}{32}r^2 + \frac{11}{384}r^4\right)/
\left(1+\frac 1 3 r^2+\frac{11}{384} r^4\right)$,
and phase distribution ${\theta_0}_k$.
The initial condition due to the presence of both rings is thus
$\psi_0 = \sqrt{{\rho_0}_1{\rho_0}_2} 
\exp{[\rmi \left({\theta_0}_1+{\theta_0}_2\right)]}$.

\section{\label{sec:topquant}Extraction of geometric and topological quantities}
Normalized total length $L/\xi$, writhe $Wr$, normalized total torsion $T$ and intrinsic twist $N$ are 
the global geometric quantities we want to monitor during reconnection. The total twist is 
given by $Tw=T+N$, and together with $Wr$ gives 
the self-linking number $SL=Wr+Tw$, a topological invariant. These quantities are well-defined 
(assuming everything sufficiently smooth) for each individual vortex ring. 

The writhe $Wr=Wr(C)$ is analytically defined by 
\begin{equation}\label{eq:Wrdef}
  Wr=\frac{1}{4\pi}\int_{C}\int_{C} \frac{\mathbf{x}-\mathbf{x}^*}
     {\left\|\mathbf{x}-\mathbf{x}^*\right\|^{3}} \cdot
     \left(\de\mathbf{x}\times\de\mathbf{r}^*\right)\ .
\end{equation}
where $C$ is the vortex centerline and $\mathbf{x}$ and $\mathbf{x}^*$ denote 
the position vectors of two points on $C$. 

The normalized total torsion $T=T(C)$ is given by 
\begin{equation}\label{eq:TTdef}
  T = \dfrac{1}{2\pi} \int_{C}  \tau(s) \de s
\end{equation}
where (from its basic definition) the local torsion $\tau(s)$, function of arc-length $s$ on $C$, 
involves third order derivatives of the position vector 
$\mathbf{x}$ of any point on $C$. 

Intrinsic twist $N=N(R)$ measures the rotation around $C$ of a reference ribbon $R$ (with  baseline 
$C$) as we move along $C$. Here $R$ has edges given by $C$ and $C'$, a second curve obtained by
translating $C$ a small 
distance $\epsilon$ (the width of $R$) along a unit normal vector $\mathbf{\hat u}$ to $C$. 
$\epsilon$ is chosen to be constant along $C$ and sufficiently small compared to the local 
radius of curvature. Clearly $R$ depends on 
the choice of $\mathbf{\hat u}$ and in absence of inflection points, this is always 
well-defined~\cite{RM92,MR92}. If $\varphi(s)$ denotes the angle between 
$\mathbf{\hat u}$ and $\mathbf{\hat n}$, we have 
\begin{equation}\label{eq:ITdef}
N=\dfrac{1}{2\pi}\int_{C}\frac{\de\varphi(s)}{\de s}\,\de s=
\dfrac{\left[ \Phi \right]_C}{2\pi}
\end{equation}
that measures the number of full rotations of the ribbon $R$, after one full turn 
along $C$.
From the definition of total twist one can show~\cite{MR92} that indeed $Tw=T+N$.

\begin{figure}[h!!!]
\centering
\includegraphics[scale=0.6]{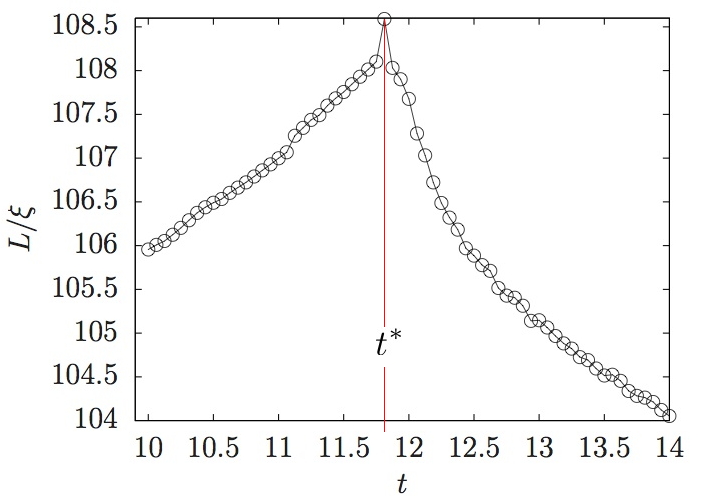}
\caption{\label{FIGURE2}
Normalized total length $L/\xi$, plotted against time $t$. The peak in $L/\xi$ 
is taken as signature of the reconnection time at $t=t^*=11.81$.}
\end{figure}

\section{\label{sec:results}Results}
Vortex centerlines are extracted from numerical data, first by isolating
the tubes whose density $\rho<0.2$, and then by looking for minima of 
$\rho/|{\boldsymbol \omega}|$ (minima of density correspond to maxima of vorticity).
Particular care has been put to the extraction of sufficiently smooth data.
Intrinsic twist is obtained from phase information. The ribbon $R$ is thus 
obtained by requiring constant phase $\theta=\bar\theta$ along $C$, and by 
setting $\epsilon=0.3$, a good compromise between visualization needs and 
misleading effects. As usual, smoothing was applied to ensure sufficient regularity.

Figure~\ref{FIGURE1} shows four snapshots of the time-evolution of interaction and reconnection 
of the quantum vortex rings (isosurfaces of $\rho=0.06$). 
Before reconnection, the two vortex rings move toward each other, bending upwards
in the region near the reconnection site, the more distant parts of the vortices remaining 
almost un-affected. The change of normalized total length $L/\xi$ of the pair of 
rings against time is used to check the reconnection process and to the detect  
reconnection time. The plot is shown in Figure~\ref{FIGURE2} for $t\in(10,14)$.
The marked peak at $t=t^*= 11.81$, after stretching, is taken as signature of the reconnection time. The maximum value $L_\mathrm{max}\approx 108.6\,\xi$ corresponds 
to about 8\% of increase with respect to the initial total 
length, given by $L_0\approx 100.5\,\xi$. For $t>t^*$ the system relaxes at a 
faster rate, confirming the time asymmetry found in earlier work~\cite{ZCBB12}.

\begin{figure}
\centering
\includegraphics[scale=0.35]{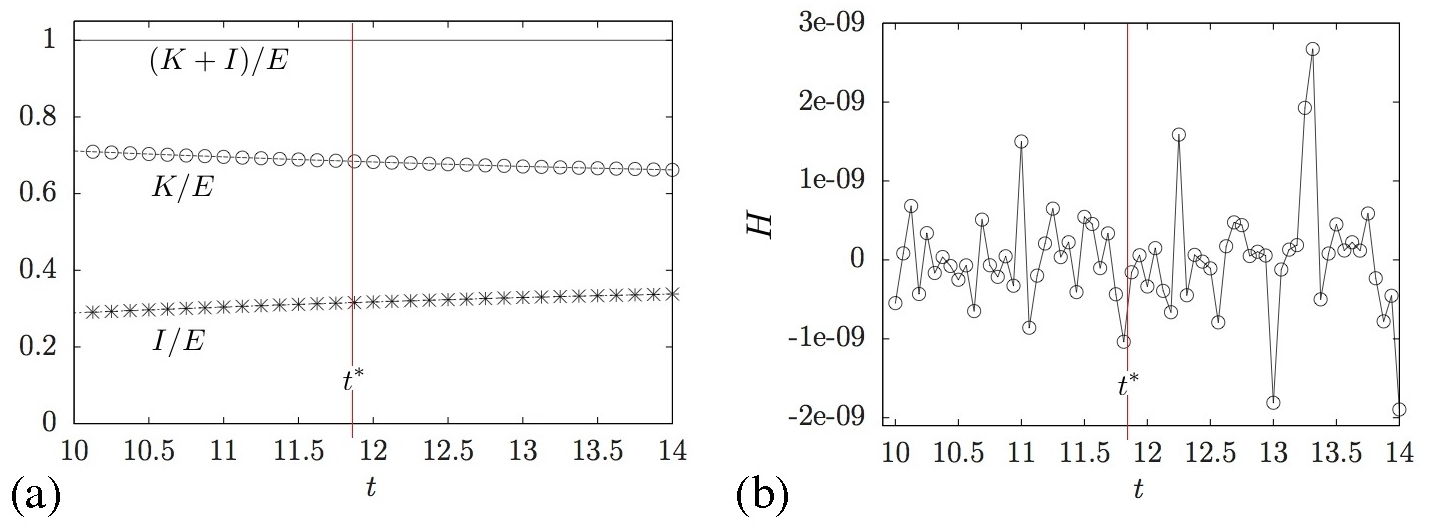}
\caption{\label{FIGURE3}
(a) Normalized total energy (hamiltonian) $(K+I)/E$, normalized kinetic energy 
$K/E$ and normalized interaction energy $I/E$ plotted against time $t$. (b) 
Kinetic helicity $H$ plotted against time $t$. The vertical line (red online) denotes 
the reconnection time $t=t^*= 11.81$.}
\end{figure}

\begin{figure}
\centering
\includegraphics[scale=0.65]{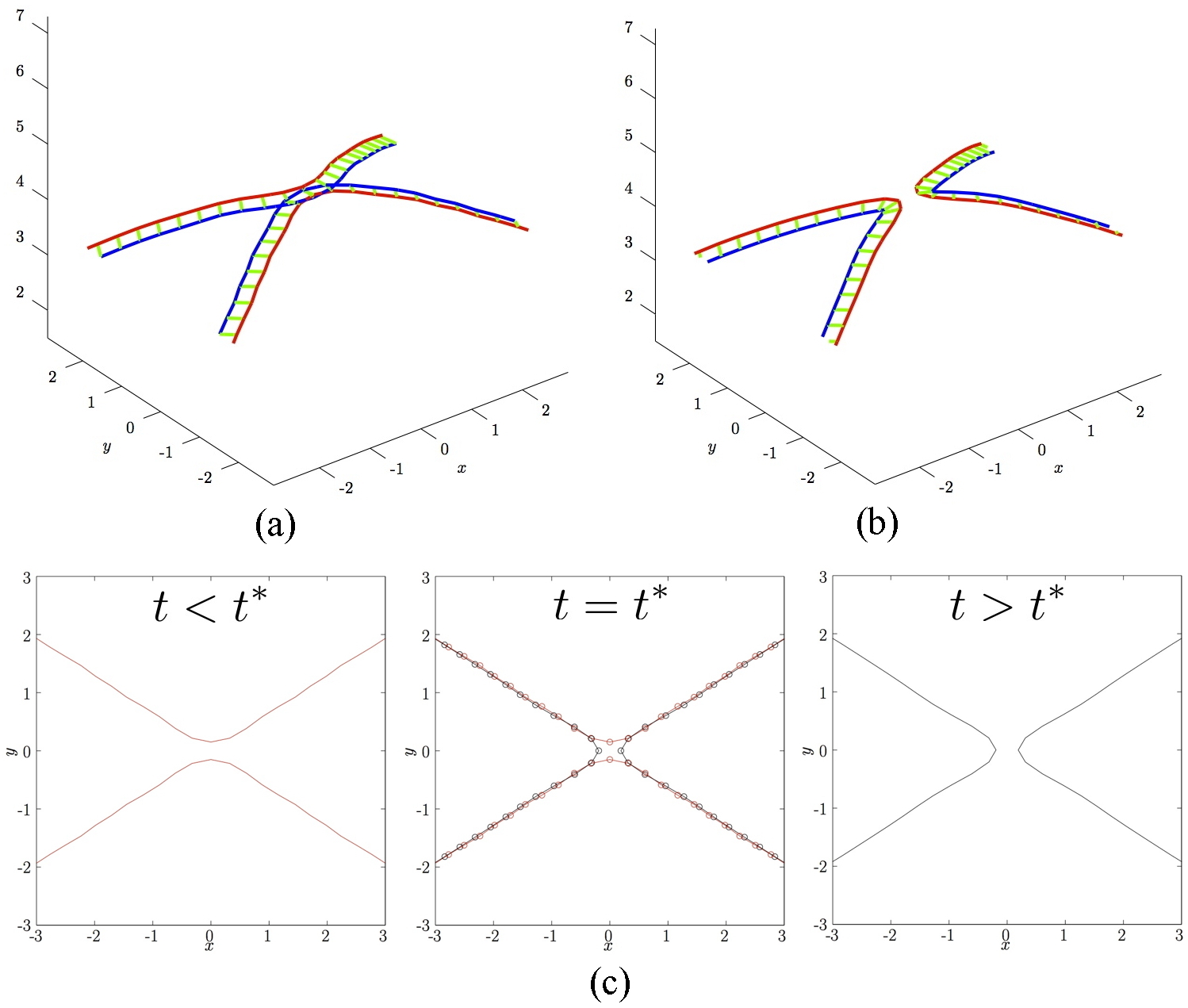}
\caption{\label{FIGURE4}
A close-up view of the vortex centerlines (red online) and reference 
ribbons (green and blue online, otherwise pale grey and darker) (a) 
immediately before reconnection at $t = 11.81$, 
and (b) immediately after reconnection at $t=11.88$. The transition is shown at maximum
numerical resolution: (c) vortex centrelines just before (red online) and after (black)
reconnection; the state in-between (at $t=t^*$) shows that reconnection is numerically 
triggered by a jump at just two nodal points (circles).}
\end{figure}

\begin{figure}
\centering
\includegraphics[scale=0.36]{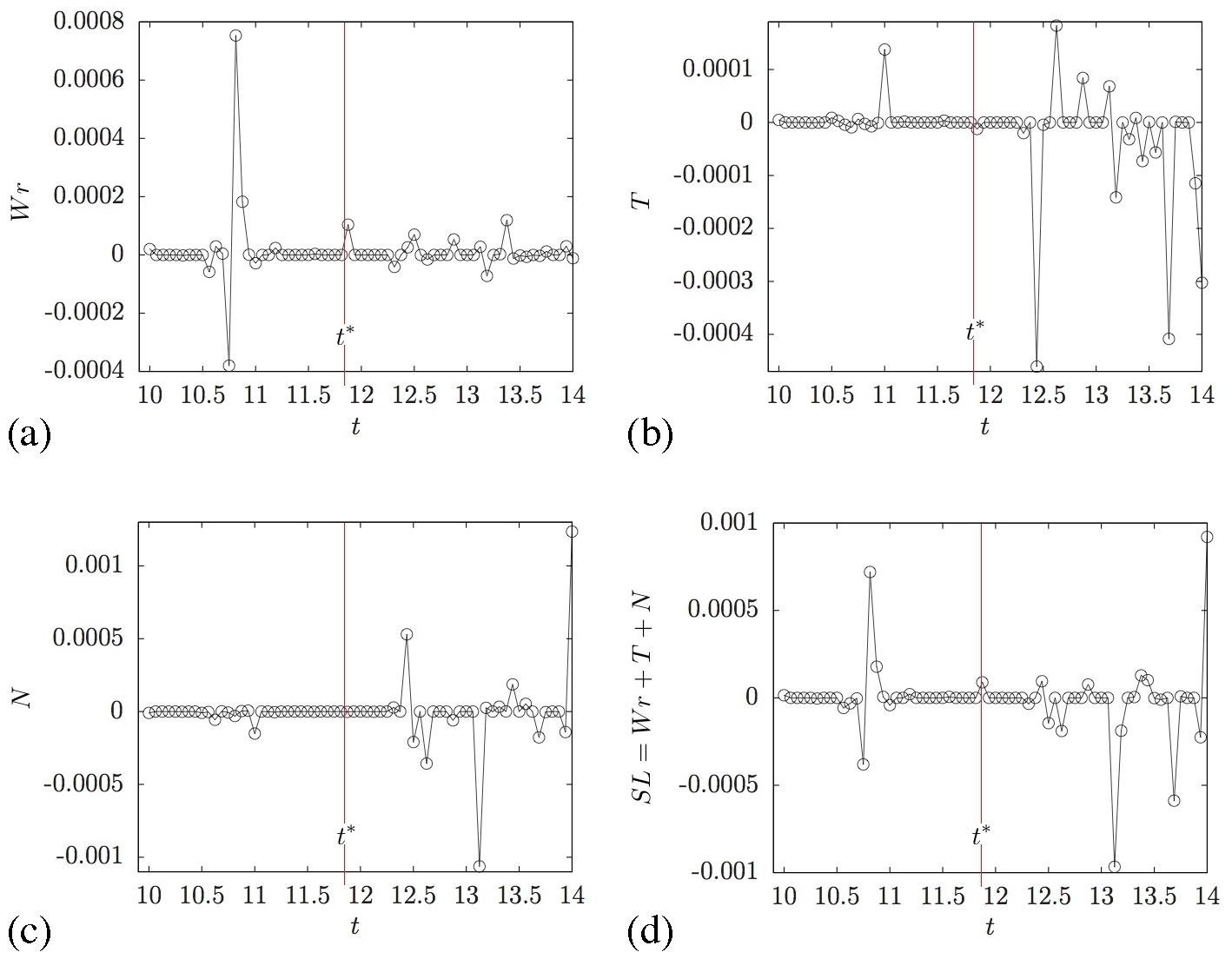}
\caption{\label{FIGURE5}
(a) Writhe $Wr$, (b) normalized total torsion $T$, (c) normalized intrinsic 
twist $N$ (with $\bar\theta\approx 50\dg$) and self-linking number $SL$ plotted 
against time $t$. The vertical line (red online) denotes the reconnection time $t=t^*= 11.81$.}
\end{figure}

As a further check, we plot  the hamiltonian ($E$) 
given by the normalized total energy $(K+I)/E$ and, separately, 
the normalized kinetic energy $K/E$ and interaction energy $I/E$, given by \eqref{eq:ham} 
(see Figure~\ref{FIGURE3}a). Kinetic helicity is computed according to eq.~\eqref{eq:Hdef}. 
As shown in Figure~\ref{FIGURE3}b its value remains bounded, i.e. 
$|H|<10^{-9}$, that is approximately zero throughout the reconnection 
process  (at these length scales the spikes of the plot in Figure~\ref{FIGURE3}b 
are essentially due to numerical noise).
A check on vortex strength confirms the conservation of $\Gamma$
before and after reconnection. A close-up view of the vortex centerlines (in red, online) 
and reference ribbons (green and blue, online) immediately before and after 
reconnection is shown in the plots of Figure~\ref{FIGURE4}a,b. The reconnection event takes 
place at a much faster timescale, well beyond numerical accuracy. To monitor as close as 
possible the topological transition, the event is represented at maximum numerical resolution by showing the 
diagrams of the discretized vortex centrelines in Figure~\ref{FIGURE5}c.
As we can see from the central diagram of Figure~\ref{FIGURE5}c 
(at $t=t^*$) the reconnecting event is 
numerically triggered by a jump at the two nodal points (circles) of closest approach, 
demonstrating that in the limit of numerical resolution reconnection involves only the mutual cancellation of two anti-parallel polygonal segments.

Finally, we examine the individual contributions to the self-linking number 
by using the independent equations \eqref{eq:Wrdef}--\eqref{eq:ITdef}. Plots 
of $Wr$, $T$, $N$ and $SL$ against time are shown in Figure~\ref{FIGURE5}.
The ribbon $R$ is found to be $\theta=\bar\theta \approx 50\dg$. Writhe and twist 
remain very small throughout the process. They are identically zero only 
at $t=0$, when the vortex rings are exactly planar tori, whereas for $t>0$ 
the vortex centerlines become gradually deformed.
Except for a few spikes, which are not related to reconnection, 
$|Wr|<10^{-4}$ and $|Tw|<2\times 10^{-4}$.
Numerical errors associated to the computation of $Tw$ are 
generally larger than those on $Wr$, because of the higher-order 
derivatives involved in the computation of the normalized total torsion 
(see Figure~\ref{FIGURE5}b) and the additional numerical noise 
associated with the computation of $N$ (see Figure~\ref{FIGURE5}c).
The numerical code has been validated by computing $Wr$ and $Tw$ of known
benchmarks, and we are confident that the reported spikes are only due 
to accumulation of numerical errors. Thus, we conclude that all plots 
of Figure~\ref{FIGURE5} show consistently $Wr=T=N=SL=0$ throughout the 
reconnection process.

\section{\label{sec:conclusions}Conclusions}
We have performed numerical simulations of the GPE, that 
resolve the fine structure of the vortex-core under anti-parallel 
reconnection of the tube strands of two colliding quantum vortex rings. 
This simple scenario provides a good benchmark for comparison with earlier works 
on direct numerical simulation of reconnecting vortex rings under Navier-Stokes 
equations~\cite{KTH89, CKL03}, 
and an ideal setup for clarifying recent contradictory 
results obtained by experiments and theoretical modeling on classical 
vortex dynamics.

Reconnection under Gross-Pitaevskii is clearly manifested by the generation of a peak in
total length, and this is taken as a marker of the reconnection event. We took extra care to monitor  
the behaviour of several geometric quantities during reconnection.  As predicted by 
geometric analysis~\cite{LRS15} writhe and total torsion are found to remain conserved, 
whereas there is no change in total intrinsic twist, all this keeping clearly 
self-linking number invariant. Self-helicity, computed independently by using 
eq.~\eqref{eq:Hdef}, is found consistently to remain conserved during reconnection. 
Since in our experiment the (Gauss) linking number $Lk$ is always zero 
(the rings remain unlinked throughout the process), 
there is no contradiction with the fact that during reconnection topology 
actually changes (as indeed happens here). The fact that self-helicity, and hence total helicity, 
remains conserved during reconnection is thus something not only new for quantum systems, but also in good agreement with the recent experimental observations 
of reconnecting vortices in water~\cite{SKP14}. The methods used here can certainly be extended to study more complex topologies 
and further work is indeed in progress to analyze and to extend this preliminary findings. 



\begin{thebibliography}{99}
\bibitem{KT1994}
  S. Kida and M. Takaoka, Annu. Rev. Fluid Mech. \textbf{26}, 169 (1994).
\bibitem{HD2011}
  A.K.M.F. Hussain and K. Duraisamy,  Phys. Fluids \textbf{23}, 021701 (2011).
\bibitem{PFL2010}
  M.S. Paoletti \etal , Physica D \textbf{239}, 1367 (2010).
\bibitem{ZCBB12}
  S. Zuccher, \etal , Phys. Fluids \textbf{24}, 125108 (2012).
\bibitem{LF96}  
  Y.T. Lau, and J.M. Finn,   Phys. Plasmas \textbf{3}, 3983 (1996).  
\bibitem{PF2000}
 E. Priest and T. Forbes, Magnetic Reconnection, Cambridge University Press (2000).
\bibitem{L99}    
  S. Lugomer, J. Fluids \& Structures \textbf{13}, 647 (1999). 
\bibitem{VCP98}
  A.V. Vologodskii \etal , J. Mol. Biol. \textbf{278}, 1 (1998).
\bibitem{VIN2008}
  W.F. Vinen, Phil. Trans. R. Soc. A \textbf{366}, 2925 (2008).
\bibitem{BSS2014}
  C.F. Barenghi \etal , Proc. Natl. Acad. Sci. USA \textbf{111}, 4647 (2104).
\bibitem{K11}
  R.M. Kerr, Phys. Rev. Lett. \textbf{106}, 224501 (2011).
\bibitem{SKP14}
  M.W. Scheeler \etal , Proc. Natl. Acad. Sci. USA \textbf{111}, 4647 (2014).
\bibitem{LRS15}
  C.E. Laing \etal , Sci. Rep. \textbf{5}, 9224 (2015).  
\bibitem{KM14}
  Y. Kimura and H.K. Moffatt, J. Fluid Mech. \textbf{751}, 329 (2014).  
\bibitem{S1992}
  P.G. Saffman, Vortex Dynamics, Cambridge University Press (1992).   
\bibitem{MOF69}
  H.K. Moffatt, J.Fluid Mech. \textbf{35}, 117 (1969).
\bibitem{RM92}
  R.L. Ricca and H.K. Moffatt, in \textit{Topological Aspects of the 
  Dynamics of Fluids and Plasmas} [H.K. Moffatt \etal\ (ed.)] p. 225,   
  Kluwer Acad. Publs. (1992).   
\bibitem{MR92}
  H.K. Moffatt and R.L. Ricca, Proc. R. Soc. Lond. A \textbf{439}, 411 (1992). 
\bibitem{Calu61}
  G. C\u alugare\u anu, Czech. Math. J. \textbf{11}, 588 (1961).   
\bibitem{White69}
  J.H. White, Am. J. Math. \textbf{91}, 693 (1969).   
\bibitem{KL1993}
  J. Koplik and H. Levine, Phys. Fluids \textbf{71}, 1375 (1993).  
\bibitem{WTZ2014}
  P.M. Walmsley \etal , Phys. Rev. Lett. 
  \textbf{113}, 125302 (2014). 
\bibitem{BER2004}
  N.G. Berloff, J. Phys. A: Math. Gen. \textbf{37}, 1617 (2004).
\bibitem{KTH89}
  S. Kida, M. Takaoka and F. Hussain, Phys. Fluids A \textbf{1}, 630 (1989). 
\bibitem{CKL03}
P Chatelain, D Kivotides and A Leonard, Phys. Rev. Lett.
  \textbf{90}, 054501 (2003). 
\end{thebibliography}
\end{document}